\documentclass[aps,11pt,nofootinbib,endfloats]{revtex4}
\usepackage{graphicx}% Include figure files
\usepackage{dcolumn}% Align table columns on decimal point
\usepackage{bm}% bold math
\usepackage{amsfonts}
\begin{document}

\date{\today}
\preprint{Brown-HET-1472}
\title{Non-Gaussianities in Multi-field Inflation}

\author{Thorsten Battefeld$^{1)}$} \email[email: ]{battefeld@physics.brown.edu}
\author{Richard Easther $^{2)}$} \email[email: ]{richard.easther@yale.edu}

\affiliation{1) Dept. of Physics, Brown University, Providence R.I.
02912, U.S.A.} 
\affiliation{2) Dept. of Physics, Yale University, New Haven CT 06520 }

\begin{abstract}
We compute the amplitude of the non-Gaussianities in  inflationary models with multiple, uncoupled scalar fields.   This calculation thus applies to all models of assisted inflation, including N-flation, where inflation is driven by multiple axion fields arising from shift symmetries in a flux stabilized string vacuum.  The non-Gaussianities are associated with nonlinear evolution of the field (and density) perturbations, characterized by the parameter  $f_{NL}$. We derive a general expression for the nonlinear parameter,  incorporating the evolution of perturbations after horizon-crossing. This is valid for arbitrary separable potentials during slow roll.    To develop an intuitive understanding of this system and to demonstrate the applicability of the formalism we examine several cases with quadratic potentials: two-field models with a wide range of mass ratios, and a general $\mathcal{N}$-field model with a narrow mass spectrum.  We uncover that $f_{NL}$ is suppressed as the number of e-foldings grows, and that this suppression is increased in models with a broad spectrum of masses.   On the other hand, we find no enhancement to $f_{NL}$ that increases with the number of fields. We thus conclude that the production of a large non-Gaussian signal in multi-field models of inflation is very unlikely as long as fields are slowly rolling and potentials are of simple, quadratic form. Finally, we compute a spectrum for the scalar spectral index that incorporates the nonlinear corrections to the fields' evolution.

\end{abstract}
\maketitle
\newpage

\tableofcontents 

%%%%%%%%%%%%%%%%%%%%%%%%%%%%%%%%%%%%
\section{Introduction}
%%%%%%%%%%%%%%%%%%%%%%%%%%%%%%%%%%%%

Multi-field models of inflation have been considered ever since the introduction of hybrid inflation \cite{Linde:1991km,Linde:1993cn,Copeland:1994vg}.   In these models, only one field typically evolves during inflation. The role of the second field is to add a potentially tachyonic direction to the potential which ends inflation   by creating  an instability in a direction orthogonal to the classical inflationary trajectory. Conversely, assisted inflation \cite{Liddle:1998jc} relies on ${\cal N}$ uncoupled fields. The fields are each  unable to generate a workable period of inflation on their own, but can evolve coherently to provide a cosmologically acceptable inflationary epoch.   

The analysis here applies to any model where inflation is driven by multiple scalar fields that do not interact directly  with one another.  This restriction is less onerous than it might appear, since assisted inflation typically only works to the model builder's advantage when the individual  fields are very weakly coupled to one another \cite{Kanti:1999vt}.   Our treatment of this problem is motivated by N-flation \cite{Dimopoulos:2005ac,Easther:2005zr} \footnote{See also \cite{Becker:2005sg} for the derivation of a closely related assisted inflation model based on multiple branes, to which our study also applies.}. This is a proposal to implement assisted inflation in the context of string theory, and the ${\cal N}$ fields arise from axion potentials associated with shift symmetries in the compact manifold.  From the model-building perspective, this   provides a mechanism for generating inflation without  invoking field(s) whose VEVs become trans-Planckian at some point during their evolution -- a point at which perturbative descriptions of any string or supergravity derived model will generically break down.  Moreover, the shift symmetries of the axions suppress couplings between the fields, a prerequisite for successful assisted inflation.   It is not yet known whether the microscopic physics of string theory will permit N-flation to naturally arise  in a realistic scenario, but it certainly represents a novel and interesting approach to string inflation.  Originally, the axion fields that drive N-flation were assumed to have identical masses \cite{Dimopoulos:2005ac}. Moreover, the sinusoidal axion potentials were approximated by Taylor expansions about their minima, retaining only the quadratic  term.  Subsequently, Easther and McAllister  showed that the mass spectrum  could be derived via random matrix arguments \cite{Easther:2005zr}, avoiding the intractable calculation  required by a direct assault on the problem.  The resulting distribution of masses   conforms to the Mar\v{c}henko-Pastur distribution, and is controlled by a single parameter $\beta$ -- the ratio between the number of axions ${\cal N}$ and the total dimension of the moduli space.  

In the limiting case where all the masses and initial field values are identical\footnote{When assisted inflation is driven by $m^2 \phi^2$ potentials  the inflationary dynamics have some ``memory'' of the initial field values, whereas other widely studied implementations of assisted inflation have an attractor solution in field space.}  the spectrum of scalar and tensor perturbations from N-flation is identical to that generated by a single massive field.   Since multi-field models of inflation generically have a richer phenomenology than single-field examples, we might hope to break this degeneracy by looking for specific signatures of multi-field evolution within N-flation. In particular, the non-Gaussianities in single field models are typically tiny \cite{Maldacena:2002vr,Acquaviva:2002ud,Creminelli:2003iq,Babich:2004gb,Seery:2005wm}, but they can be larger in general multi-field models, so computing them is an important step towards developing a full understanding of N-flation \footnote{An additional source of Non-Gaussianity is the era of reheating, which we do not consider in the following; see eg. \cite{Barnaby:2006km,Enqvist:2004ey} and references therein for relevant literature.}.   Several studies have argued that the non-Gaussianities from assisted inflation are small \cite{Vernizzi:2006ve,Seery:2005gb,Kim:2006te}, but these calculations contain significant simplifying assumptions. For example,  only two fields are considered in \cite{Vernizzi:2006ve}, and  \cite{Seery:2005gb,Kim:2006te} use the horizon crossing approximation.  As a result, a general expression for the non-Gaussianities, described by $f_{NL}$, is lacking. To be more precise, terms  are usually neglected  that incorporate the evolution of perturbations once they cross the horizon and the effects of isocurvature modes (both of which are always possible in multi-field models) -- see for example \cite{Kim:2006te}.

We derive this desired general expression within the $\delta N$-formalism, first proposed by Starobinsky \cite{Starobinski} and extended by Sasaki and Stewart \cite{Sasaki:1995aw} among others \cite{Lyth:2005fi,Vernizzi:2006ve}. The only assumptions we make are that the potential is separable and that the slow roll approximation is valid. Since the general expression contains a part that is not immediately slow roll suppressed, we will focus our efforts on this term.  To illustrate the application of the formalism and to build up intuition we consider several specific cases with quadratic potentials: first, an exactly solvable two-field model where the ratio of the squared masses is two; second, generic two-field models where the ratio of the squared masses is less than five; third, the generic multi-field case with a narrow mass spectrum.

In the two-field cases, we find an unexpected suppression of $f_{NL}$ by the volume expansion rate, expressed in terms of the number of e-foldings. The exponent of the rate is given by twice the square of the ratio of the heavier mass to the lighter one. Based on this result, we focus our attention on narrow mass spectra, for which we compute the nonlinear parameter up to second order in the width of the spectrum (properly defined in Sec.~\ref{case3}). As expected, we encounter a suppression factor that scales as the square of the number of e-foldings.   However, we do not find any enhancement that scales with the total number of fields.     Possible exceptions to this argument include models where one or more fields violate slow roll, or where there are significant couplings between the fields. We intend to address these issues in future work. However, at present we find no evidence that the non-Gaussianities generated by assisted inflation modes -- including N-flation -- are enhanced relative to those of their single field analogs. 

The article is structured as follows: first, we derive the general expression for $f_{NL}$ in Sec.~\ref{nongaussianities}, followed by a discussion where we reduce this general result to recover the specific cases already encountered in the literature. Thereafter, we discuss three specific cases, two-field models in Sec.~\ref{case1} and \ref{case2} and the multi-field model in Sec.~\ref{case3}, again followed by a discussion. Last but not least, we derive the general expression of the scalar spectral index incorporation the evolution of perturbations once they cross the horizon in Sec.~\ref{scind}, but we postpone specific case studies to a follow-up publication. We conclude in Sec.~\ref{sec:con}.

\section{Non-Gaussianities in Multi-field Inflation \label{sec1}}
We are interested in evaluating non-Gaussianities in models of inflation with multiple scalar fields, in the hope of finding possible experimental signals that will break the degeneracy between multi-field models and common single field models.\footnote{See  \cite{Seery:2005gb,Lyth:2005fi,Alabidi:2005qi,Rigopoulos:2005us} for a sample of the recent literature on multiple field models and e.g. \cite{Maldacena:2002vr,Acquaviva:2002ud,Chen:2006nt} for a discussion of single field models.}   We shall use the $\delta N$-formalism to compute $f_{NL}$, which characterizes non-Gaussianities. This formalism was proposed by Starobinsky \cite{Starobinski} and further developed by Sasaki and Stewart in \cite{Sasaki:1995aw}, and others in \cite{Seery:2005gb,Lyth:2005fi,Vernizzi:2006ve}.  In this approach one relates the perturbation of the volume expansion rate $\delta N$ to the curvature perturbation $\zeta$ on large scales, which is possible if the initial hypersurface is flat and the final one is a uniform density hypersurface \cite{Sasaki:1995aw}. Note that $\zeta$ is conserved on large scales in simple models, even beyond linear order \cite{Lyth:2003im,Rigopoulos:2003ak} \footnote{The separate Universe formalism put forward by Rigopoulos and Shellard in e.g. \cite{Rigopoulos:2003ak} is equivalent to the $\delta N$-formalism.}. Given this relationship between the curvature perturbation and the volume expansion rate, one can evaluate the momentum independent pieces of non-linear parameters, which are related to higher order correlation functions, in terms of the change in $N$ during the evolution of the Universe, see e.g. \cite{Byrnes:2006vq}. 

Our treatment will parallel that of Vernizzi and Wands \cite{Vernizzi:2006ve}, who computed the general expression for $f_{NL}$, but restricted themselves to two-field models when they came to compute the non-Gaussianities. 

To begin, consider the action for $\mathcal{N}$ scalar fields,
\begin{eqnarray}
S&=&\frac{m_p^2}{2}\int d^4x \sqrt{-g} \left(\frac{1}{2}\sum_{i=1}^{\mathcal{N}}\partial^\mu\varphi_i\partial_\mu\varphi_i+W(\varphi_1,\varphi_2,...)\right) 
\end{eqnarray}
which we assume to be responsible for driving an inflationary phase. The unperturbed volume expansion rate from an initial flat hypersurface at $t^*$ to a final uniform density hypersurface at $t^c$ is given by
\begin{eqnarray}
N(t_c,t_*)\equiv\int_*^c H dt \,,\label{nofh}
\end{eqnarray}
where $H$ is the Hubble parameter.
The nonlinear parameter $f_{NL}$ can be related to the derivatives of the expansion rate $N$ with respect to the field values $\varphi_i(t^*)\equiv\varphi_i^*$. This computation, starting from the three point correlation function, was performed in \cite{Seery:2005gb,Vernizzi:2006ve} resulting in
\begin{eqnarray}
-\frac{6}{5}f_{NL}&=&\frac{r}{16}(1+f)+\frac{\sum_{i,j=1}^{\mathcal N} N_{,i}N_{,j}N_{,ij}}{\left(\sum_{i=1}^{\mathcal N} N^2_{,i}\right)^2}\,, \label{fnl}\\
&\equiv&\frac{r}{16}(1+f)-\frac{6}{5}f_{NL}^{(4)}\,,
\end{eqnarray}
where the short hand notation
\begin{eqnarray}
N_{,i}&\equiv& \frac{\partial N}{\partial \varphi_i^*}\,,\\
N_{,ij}&\equiv& \frac{\partial^2 N}{\partial \varphi_i^*\partial \varphi_j^*}\,,
\end{eqnarray}
was used (we refer the reader to \cite{Seery:2005gb,Vernizzi:2006ve} for details). The first term in (\ref{fnl}) is small. On geometrical grounds, we know $0\leq f\leq 5/6$ \cite{Maldacena:2002vr,Vernizzi:2006ve}, while $r$ is the usual tensor:scalar ratio.\footnote{The quantity $f$ incorporates the dependence of the three point correlation function on the shape of the momentum triangle; the maximal value results for an equilateral triangle and the minimal one if two sides are much longer than the third \cite{Maldacena:2002vr}.} The observational upper limit on this quantity depends on the priors used in the fitting process, but we can reliably conclude that $r/16 < 0.1$ \cite{Spergel:2006hy}.  Observationally, it is very unlikely we will ever detect non-Gaussianities unless $f_{NL} > 1$. Henceforth, we focus on the second term in (\ref{fnl}),  to determine under what conditions non-Gaussianities could become large. Currently, the best observational bound on $f_{NL}$ is drawn from  the WMAP3 data \cite{Spergel:2006hy}: $-54 < f_{NL} < 114$.
Recently, \cite{Kim:2006te} Kim and Liddle derived the estimate $(6/5)f_{NL}^{(4)}\leq r/16$ by ignoring the evolution of perturbations once they cross the horizon, and constraining the potential $W$ to be the sum of monomials in $\varphi_i$.  Conversely,  \cite{Vernizzi:2006ve} is restricted to the two-field case but includes effects from evolution that occurs after horizon crossing, and arrives at a similar conclusion. 
Since scalar perturbations need not freeze out after horizon crossing in multi-field models, we extend \cite{Vernizzi:2006ve,Kim:2006te} to compute $f_{NL}^{(4)}$ for ${\cal N}$ fields without assuming any freeze-out. Our principal assumption is that  the slow roll approximation is valid for all fields -- in practice, individual fields can cease to be critically damped well before inflation comes to an end, depending on their masses and initial values.  Since we retain any possible super-horizon evolution, our analysis follows a similar path to \cite{Vernizzi:2006ve}.

\subsection{Slow Roll Dynamics}
In the following, we will set the reduced Planck mass $m_p=(8\pi G)^{-1/2}\equiv 1$ for notational simplicity. Since we are interested in assisted inflation, we  can ignore  cross coupling terms between the scalar fields. That is, we assume
\begin{eqnarray}
W(\varphi_1,\varphi_2,...)=\sum_{i=1}^{\mathcal N}V_i(\varphi_i)\,,
\end{eqnarray}
but we keep the form of the potentials $V_i(\varphi_i)$ general. The equations of motion for the fields are
\begin{eqnarray}
\ddot{\varphi}_i+3H\dot{\varphi}_i+V_i^\prime=0\,,
\end{eqnarray}
where we set $V_{i}^\prime=W_{,i}\equiv\partial V_i/\partial\varphi_i$. The Friedman equations read
\begin{eqnarray}
H^2&=&\frac{1}{3}\left(W+\sum_{i=1}^{\mathcal N}\frac{1}{2}\dot{\varphi}_i^2\right)\,,\\
\dot{H}&=&-\sum_{i=1}^{\mathcal N}\frac{1}{2}\dot{\varphi}_i^2\,.
\end{eqnarray}
Slow roll inflation occurs if the slow roll parameters, defined as\footnote{We are using the potential slow roll formalism here. For a more general treatment of the slow roll expansion with multiple fields, see \cite{Easther:2005nh}.}
\begin{eqnarray}
\varepsilon_i\equiv\frac{1}{2}\frac{V_{i}^{\prime 2}}{W^2}\hspace{0.5cm},\hspace{0.5cm}\eta_i\equiv\frac{V_{i}^{\prime\prime}}{W}\,, \label{srparameters}
\end{eqnarray}
are small ($\varepsilon_i\ll 1$, $\eta_i\ll 1$), and    
\begin{eqnarray}
\varepsilon\equiv\sum_{i=1}^{\mathcal N}\varepsilon_i\ll1 \, .
\end{eqnarray}
 In this case the dynamics is governed by
\begin{eqnarray}
3H\dot{\varphi}_i&\approx&-V_i^\prime\,,\label{dyn1}\\
3H^2&\approx& W\,.\label{dyn2}
\end{eqnarray}
For simplicity, we assume $V_i^\prime>0$ from here on, so that we can replace the derivatives by the slow roll parameters,  $V_i^\prime=W\sqrt{2\varepsilon_i}$.

During slow roll inflation we can write the number of e-foldings (\ref{nofh}) as \cite{Lyth:1998xn}
\begin{eqnarray}
N(t_c,t_*)=-\int_*^c \sum_{i=1}^{\mathcal N}\frac{V_i}{V_i^\prime}d\varphi_i \,.\label{defN}
\end{eqnarray}
Furthermore, there are ${\mathcal N}-1$ integrals of motion, for example the set 
\begin{eqnarray}
C_i\equiv-\int\frac{d\varphi_i}{V_i^\prime}+\int\frac{d\varphi_{i+1}}{V_{i+1}^\prime}\,,\label{IntOM}
\end{eqnarray}
for $i=1\,...\,{\mathcal N}-1$. These $C_i$ can be used to discriminate between different trajectories in field space and they will become quite handy in the next subsection when we evaluate $N_{,i}$ and $N_{,ij}$. 

\subsection{Non-Gaussianities \label{nongaussianities}}
We will now evaluate the derivatives of the volume expansion rate with respect to the fields, which are needed in order to evaluate the nonlinear parameter. First, write down the total differential of $N$  by using its definition (\ref{defN}),
\begin{eqnarray}
dN=\sum_{j=1}^{\mathcal N}\left[\left(\frac{V_j}{V_j^\prime}\right)_*-\sum_{i=1}^{\mathcal N}\frac{\partial \varphi_i^c}{\partial \varphi_j^*}\left(\frac{V_i}{V_i^\prime}\right)_c\right]d\varphi_j^*\,.\label{dN}
\end{eqnarray}
Furthermore, using the integrals of motion $C_i$ from (\ref{IntOM}), we can relate $d\varphi_i^c$ and $d\varphi_i^*$ via
\begin{eqnarray}
d\varphi_j^c=\sum_{i=1}^{\mathcal N-1}\frac{\partial \varphi_j^c}{\partial C_i}\left(\sum_{k=1}^{\mathcal N}\frac{\partial C_i}{\partial \varphi_k^*}d\varphi_k^*\right)\,, \label{dvarphi}
\end{eqnarray}
where 
\begin{eqnarray}
\nonumber \frac{\partial C_i}{\partial\varphi_k^*}&=&\frac{1}{(V_k^\prime)^*}(\delta_{ik-1}-\delta_{ik})\\
&=&\frac{1}{\sqrt{2\varepsilon_k^*}W^*}(\delta_{ik-1}-\delta_{ik})\,. \label{diffC}
\end{eqnarray}

Next, we would like to make use of the $\mathcal{N}-1$ integrals of motion to eliminate $\partial \varphi_j^c/\partial C_i$ in favor of $\partial \varphi_1^c/\partial C_i$, which can then be used in the condition $\rho=\mbox{const}$ at $t_c$, the time at which we want to evaluate the non-Gaussianities. To accomplish this, consider
\begin{eqnarray}
\tilde{C_i}&\equiv&\sum_{j=1}^{i-1}C_j\\
&=&-\int\frac{d\varphi_1}{V_1^\prime}+\int\frac{d\varphi_i}{V_i^\prime}\,.
\end{eqnarray}
Differentiating this with respect to $C_k$ yields
\begin{eqnarray}
\frac{\partial \tilde{C}_i}{\partial C_k}=-\frac{\partial\varphi_1^c}{\partial C_k}\frac{1}{(V_1^\prime)_c}+\frac{\partial \varphi_i^c}{\partial C_k}\frac{1}{(V_i^\prime)_c}\,,
\end{eqnarray}
which can be solved to give 
\begin{eqnarray}
\frac{\partial \varphi_i^c}{\partial C_k}=\left(\frac{V_i^\prime}{V_1^\prime}\right)_c\frac{\partial \varphi_1^c}{\partial C_k}+(V_i^\prime)_c\Theta_{ki} \,,
\end{eqnarray}
where we introduced 
\begin{eqnarray}
\Theta_{ki}\equiv\Bigg\{\begin{array}{l}
1, \mbox{ if } k\leq i-1\\
0, \mbox{ if }  k> i-1\,.
\end{array}
\end{eqnarray}
If we plug this into the derivative (with respect to $C_l$) of the $\rho=\mbox{const}$ condition,  
\begin{eqnarray}
0= \sum_{i=1}^{\mathcal N}(V_i^\prime)_c\frac{\partial\varphi_i^c}{\partial C_k}\,,
\end{eqnarray}
after some algebra we arrive at 
\begin{eqnarray}
\frac{\partial \varphi_i^c}{\partial C_k}&=&-\left(V_i^\prime\frac{\sum_{j=k+1}^{\mathcal N} V_j^{\prime^2}}{\sum_{j=1}^{\mathcal N}V_j^{\prime 2}}\right)_c+\Theta_{ki}(V_i^\prime)_c\\
&=&-W_c\frac{\sqrt{2\varepsilon_i^c}}{\varepsilon^c}\left(\sum_{j=k+1}^{\mathcal N}\varepsilon_j^c-\Theta_{ki}\varepsilon^c\right)\,, \label{whatever}
\end{eqnarray}
where we used the definition of the slow roll parameters (\ref{srparameters}) in the last step. Using (\ref{whatever}) and (\ref{diffC}) in (\ref{dvarphi}) we end up with
\begin{eqnarray}
\frac{\partial\varphi_j^c}{\partial\varphi_k^*}=-\frac{W_c}{W_*}\sqrt{\frac{\varepsilon_j^c}{\varepsilon_k^*}}\left(\frac{\varepsilon_k^c}{\varepsilon^c}-\delta_{kj}\right)\,.
\end{eqnarray}
We are now ready to compute the derivatives of the expansion rate from (\ref{dN}), which reduce to
\begin{eqnarray}
\frac{\partial N}{\partial \varphi_k^*}=\frac{1}{\sqrt{2\varepsilon_k^*}}\frac{V_k^*+Z_k^c}{W_*}\,, \label{N,k}
\end{eqnarray}
where we introduced
\begin{eqnarray}
Z_k^c&\equiv&\frac{1}{\varepsilon^c}\sum_{i=1}^{\mathcal N}V_i^c(\varepsilon_k^c-\varepsilon^c \delta_{ki})\\
&=&W^c\frac{\varepsilon_k^c}{\varepsilon^c}-V_k^c\,.
\end{eqnarray}
 After some algebra, the second derivative becomes
\begin{eqnarray}
\frac{\partial^2 N}{\partial \varphi_k^*\partial \varphi_l^*}=\delta_{k,l}\left(1-\frac{\eta_l^*}{2\varepsilon_l^*}\frac{V_l^*+Z_l^c}{W_*}\right)+\frac{1}{\sqrt{2\varepsilon_l^*}W_*}\frac{\partial Z_l^c}{\partial\varphi_k^*}\,,\label{N,kl}
\end{eqnarray}
with
\begin{eqnarray}
\frac{\partial Z_l^c}{\partial\varphi_k^*}&=&-\frac{W^2_c}{W_*}\sqrt{\frac{2}{\varepsilon_k^*}}\left[\sum_{j=1}^{\mathcal N}\varepsilon_j\left(\frac{\varepsilon_l}{\varepsilon}-\delta_{lj}\right)\left(\frac{\varepsilon_k}{\varepsilon}-\delta_{kj}\right)\left(1-\frac{\eta_j}{\varepsilon}\right)\right]_c \label{finalZ}\\
&\equiv&\sqrt{\frac{2}{\varepsilon_k^*}}W_*\mathcal{A}_{lk}\,. \label{defA}
\end{eqnarray}
Note that $\mathcal{A}_{lk}$ is a symmetric matrix. Given that, we can now write down the general expression for $f_{NL}^{(4)}$ as
\begin{eqnarray}
-\frac{6}{5}f_{NL}^{(4)}=2\frac{\sum_{k=1}^{\mathcal N}\frac{u_k^2}{\varepsilon_k^*}\left(1-\eta_k^*\frac{u_k}{2\epsilon_k^*}\right)+\sum_{k,l=1}^{\mathcal N}\frac{u_ku_l}{\varepsilon_k^*\varepsilon_l^*}\mathcal{A}_{lk}}{\left(\sum_{k=1}^{\mathcal N}\frac{u_k^2}{\varepsilon_k^*}\right)^2}\,, \label{f_NL}
\end{eqnarray}
where $u_k$ is given by
\begin{eqnarray}
u_k&\equiv&\frac{V_k^*+Z_k^c}{W_*}\\
&=&\frac{\Delta V_k}{W^*}+\frac{W^c}{W^*}\frac{\varepsilon_k^c}{\varepsilon^c}\,,\label{defuk}
\end{eqnarray}
with $\Delta V_k\equiv V_k^*-V_k^c>0$.

Equation (\ref{f_NL}) is the desired nonlinear parameter characterizing non-Gaussianities and our first main result. The factor $Z_c$ incorporates the evolution of perturbations after they cross the horizon until the time $t_c$ -- it is this factor that was neglected in \cite{Kim:2006te}. Before we discuss (\ref{f_NL}) in the next subsection, it is worthwhile to reiterate our assumptions. First of all, we assume slow roll inflation throughout. This is a tricky assumption in inflationary multi-field models, since the heavier fields can leave slow roll while inflation still continues, see e.g. \cite{Easther:2005zr}.  In this case, computing the contribution of these fields onto $f_{NL}$ would require us to go beyond the formalism used here.  Next, we assumed a separable potential, neglecting cross coupling terms between the fields, but otherwise we kept the form of the potential general. Neglecting cross couplings is reasonable when examining assisted inflation models like N-flation, since these models require that fields do not interact to any significant degree. Last but not least, some minor technical assumptions are introduced via our use of the $\delta N$-formalism, such as the reduction of four-point functions to a product of two-point functions by means of Wick's theorem -- we refer the reader to \cite{Starobinski,Sasaki:1995aw,Lyth:2005fi,Vernizzi:2006ve} for the derivation of the $\delta N$-formalism.

\subsection{Discussion}

Our main result is the nonlinear parameter given by (\ref{f_NL}) together with (\ref{defA}). In the case of two fields, this reduces to the case examined by Vernizzi and Wands \cite{Vernizzi:2006ve}. To be specific, with $\mathcal{N}=2$ equation (\ref{defA}) becomes
\begin{eqnarray}
\mathcal{A}_{11}=\mathcal{A}_{22}=-\mathcal{A}_{12}=-\mathcal{A}_{21}=\mathcal{A}\,,
\end{eqnarray}
where we introduced
\begin{eqnarray}
\mathcal{A}\equiv-\frac{W_c^2}{W_*^2}\frac{\varepsilon_1^c\varepsilon_2^c}{\varepsilon_c}\left(1-\frac{\eta_1^c\varepsilon_2^c+\eta_2^c\varepsilon_1^c}{\varepsilon^c}\right)\,.\label{defA2field}
\end{eqnarray}
As a consequence, (\ref{f_NL}) becomes
\begin{eqnarray}
-\frac{6}{5}f_{NL}^{(4)}=2\frac{\frac{u_1^2}{\varepsilon_1^*}\left(1-\frac{\eta_1^*}{2\varepsilon_1^*}u_1\right)+\frac{u_2^2}{\varepsilon_2^*}\left(1-\frac{\eta_2^*}{2\varepsilon_2^*}u_2\right)+\left(\frac{u_1}{\varepsilon_1^*}-\frac{u_2}{\varepsilon_2^*}\right)^2\mathcal{A}}{\left(\frac{u_1^2}{\varepsilon_1^*}+\frac{u_2^2}{\varepsilon_2^*}\right)^2}\,, \label{twofieldf}
\end{eqnarray}
which is identical to (68) of \cite{Vernizzi:2006ve}. 

On the other hand, if $Z_k^c \approx 0$ we recover the simple expression of \cite{Kim:2006te} for $\mathcal{N}$ scalar fields, which resulted in a small contribution of $f_{NL}^{(4)}$ to the total nonlinear parameter.

Consequently, a large non Gaussian signal could only arise if  $Z_k^c$ becomes large. A closer look at (\ref{f_NL}) reveals that only the last term in the numerator is not immediately slow roll suppressed, that is the term proportional to the $\mathcal{A}$-matrix elements defined in (\ref{defA}).  Therefore, let us  look at this matrix and ask under what conditions its contribution could become large.

First, note that the sum over a column or a row of the $\mathcal{A}$-matrix vanishes, since
\begin{eqnarray}
\nonumber \sum_{l=1}^{\mathcal{N}}\mathcal{A}_{kl}&=&-\frac{W_c^2}{W_*^2}\left[\sum_{l,j=1}^{\mathcal{N}}\varepsilon_j\left(\frac{\varepsilon_{l}}{\varepsilon}-\delta_{lj}\right)\left(\frac{\varepsilon_{k}}{\varepsilon}-\delta_{kj}\right)\left(1-\frac{\eta_j}{\varepsilon}\right)\right]_c\\
\nonumber &=&-\frac{W_c^2}{W_*^2}\left[\sum_{j=1}^{\mathcal{N}}\varepsilon_j\left(\frac{\varepsilon_{k}}{\varepsilon}-\delta_{kj}\right)\left(1-\frac{\eta_j}{\varepsilon}\right)\sum_{l=1}^{\mathcal{N}}\left(\frac{\varepsilon_{l}}{\varepsilon}-\delta_{lj}\right)\right]_c\\
&=& 0\,,\label{rowA}
\end{eqnarray}
because $\sum \varepsilon_l=\varepsilon$ so that the last sum is identical to zero. Based on this fact one can immediately see that the contribution to the $\mathcal{A}$-matrix vanishes if we deal with $\mathcal {N}$ identical fields (as expected). But even in the generic case, we should expect some cancellation.

Next, let us go back to the definition of $u_k$ in (\ref{defuk}): if $\Delta V_k \ll W^c\varepsilon^c_k/\varepsilon^c$, we can approximate 
\begin{eqnarray}
u_k\approx \frac{W^c}{W^*}\frac{\varepsilon_k^c}{\varepsilon^c}\,.
\end{eqnarray}
Using this  in (\ref{defA}), we can write the last term in (\ref{f_NL}) as
\begin{eqnarray}
\frac{\sum_{k,l=1}^{\mathcal N}\frac{u_ku_l}{\varepsilon_k^*\varepsilon_l^*}\mathcal{A}_{lk}}{\left(\sum_{k=1}^{\mathcal N}\frac{u_k^2}{\varepsilon_k^*}\right)^2}
\approx-\left(\frac{\varepsilon^c}{\tilde{\varepsilon}_{1}}\right)^2\left[\tilde{\varepsilon}_2-\frac{\tilde{\varepsilon}_1^2}{\varepsilon^c}\left(1+\frac{\tilde{\eta}_0}{\varepsilon^c}\right)+2\tilde{\varepsilon}_1\frac{\tilde{\eta}_1}{\varepsilon^c}-\tilde{\eta}_2\right]\,, \label{approxsums}
\end{eqnarray}
where we introduced
\begin{eqnarray}
\tilde{\varepsilon}_1\equiv\sum_{k=1}^{\mathcal{N}}\frac{(\varepsilon_k^c)^2}{\varepsilon_k^*}\,\,\, , \,\,\, 
\tilde{\varepsilon}_2\equiv\sum_{k=1}^{\mathcal{N}}\frac{(\varepsilon_k^c)^3}{(\varepsilon_k^*)^2}\,,
\end{eqnarray}
and 
\begin{eqnarray}
\tilde{\eta}_0\equiv\sum_{k=1}^{\mathcal{N}}\frac{\varepsilon_k^c}{\varepsilon^c}\eta_k^c\,\,\, , \,\,\,
\tilde{\eta}_1\equiv\sum_{k=1}^{\mathcal{N}}\frac{(\varepsilon_k^c)^2}{\varepsilon^c\varepsilon_k^*}\eta_k^c\,\,\, , \,\,\,
\tilde{\eta}_2\equiv\sum_{k=1}^{\mathcal{N}}\frac{(\varepsilon_k^c)^3}{\varepsilon^c(\varepsilon_k^*)^2}\eta_k^c\,.
\end{eqnarray}

As an application, let us have a look at the  ``horizon crossing limit'' $t_c\rightarrow t_*$, which corresponds to purely adiabatic perturbations. In this limit $\tilde{\varepsilon}_i=\varepsilon^*=\varepsilon^c$ and $\tilde{\eta}_i=\sum\varepsilon_k^*\eta_k^*/\varepsilon^*$, so that (\ref{approxsums}) vanishes identically. One can see this from (\ref{f_NL}), since the prefactor in front of the $\mathcal{A}$-matrix elements becomes independent of $k$, so that a sum over the matrix elements  is identically zero due to (\ref{rowA}). The second term in (\ref{f_NL}) is sensitive to the evolution of modes after horizon crossing. This evolution and thus the non-Gaussianity is closely correlated with the presence of isocurvature perturbations. These modes generically occur in multi-field inflationary models whenever light degrees of freedom transverse to the adiabatic direction are present, but it is difficult to transfer them to the adiabatic mode during \emph{slow roll inflation}: in order to transfer isocurvature modes effectively, the trajectory in field space should be sharply curved, which might occur when a field leaves slow roll.\footnote{In N-flation, this happens frequently during inflation, since the heavier a field is the earlier it will leave slow roll \cite{Easther:2005zr}.} However, this process can not be described properly with the formalism at hand, since the slow roll conditions are necessarily violated.  In addition, we can generate isocurvature modes at the end of inflation if the different fields decay into different particle species.  

We now look  at another interesting case  for which our formalism is valid, and which might still permit non-Gaussianities, even though we do not expect them to be large. Assume the time $t_c$ lies towards the end of inflation, just before slow roll ends. Even if $\tilde{\varepsilon}_1\,,\,\tilde{\varepsilon}_2\,,\,\varepsilon^c \ll1$ we could still get a non-Gaussianity of order one if the $\eta_i$ are reasonably large. First of all, it seems possible to cancel the suppression due to $\varepsilon^c_k/\varepsilon^c$ in $\tilde{\eta}_i$, which is of order $\mathcal{O}(1/\mathcal{N})$: since we have $\mathcal{N}$ summands, we can naively expect an enhancement of order $\mathcal{O}(\mathcal{N})$ that cancels out the previous suppression. Next, if we are near the end of inflation, we have $\varepsilon_k^c\gg\varepsilon^*_k$, so that the ratio $\varepsilon_k^c/\varepsilon^*_k$ becomes large. However, this enhancement factor in $\tilde{\eta}_1$ and $\tilde{\eta}_2$ gets compensated by the prefactor.  Thus, we might naively expect $f_{NL}^{(4)}=\mathcal{O}(\tilde{\eta}_0)$ towards the end of inflation, which can be of order one.

However, we have to be more careful here: we already saw in the case of $\mathcal{N}$ identical fields that cancellations occur in the sum, due to the symmetries of the $\mathcal{A}$-matrix. In the next section we will examine a few specific models in order to develop intuition as to whether non-Gaussianities can become large. It will turn out that the naive argument above is misleading, since the sum over the $\mathcal{A}$-matrix generically leads to terms that are slow roll suppressed. Hence, it seems very hard to produce a significant non-Gaussian signal in assisted inflation models which are well-described by slow-roll.

\section{Case Studies}
In section \ref{nongaussianities} we derived the general expression for the nonlinear parameter, (\ref{f_NL}). The only term which was not obviously small during slow roll is
\begin{eqnarray}
F\equiv\frac{\sum_{k,l=1}^{\mathcal N}\frac{u_ku_l}{\varepsilon_k^*\varepsilon_l^*}\mathcal{A}_{kl}}{\left(\sum_{k=1}^{\mathcal N}\frac{u_k^2}{\varepsilon_k^*}\right)^2}\,. \label{sum}
\end{eqnarray}
Consider this expression for a few specific models: first, an ``exact'' solvable two-field model with $m_2^2=2m_1^2$; second, approximate two-field models with $m_2^2=\alpha m_1^2$ and $1\leq\alpha\leq5$; third, $\mathcal{N}$-field models with $m_1^2/m_i^2 \equiv \mu_i^2 \equiv 1-\delta_i$ where $\delta_i\ll 1$.

The first step consists always of evaluating the field values at $t_*$ and $t_c$, which we denote by $\varphi^{*}_i$ and $\varphi^c_i$. To do so, we need $2\mathcal{N}$ conditions, which are given by: $\mathcal{N}-1$ dynamical relations between the fields, $\mathcal{N}-1$ initial conditions, one condition from the requirement that $t_*$ be $N$ e-foldings before $t_c$, and one condition from the requirement that slow roll ends for at least one field at $t_c$. Thereafter, we need to evaluate the slow roll parameters at $t_*$ and $t_c$, which in turn enable us to compute $F$.

Consequently, let us first look at the dynamics of multi-field inflationary models with
\begin{eqnarray}
W&=&\sum_{i=1}^{\mathcal{N}}V_i\\
&=&\frac{1}{2}\sum_{i=1}^{\mathcal{N}}m_i^2\varphi_i^2\,,
\end{eqnarray}   
where we ordered the fields such that $m_i>m_j$ if $i>j$. The field equations (\ref{dyn1}) during slow roll yield
\begin{eqnarray}
\dot{\varphi}_i=-\frac{m_i^2\varphi_i}{\sqrt{3}W}\,,
\end{eqnarray} 
where we also used the Friedman equation (\ref{dyn2}). Dividing equations such that $W$ drops out and integrating yields the $\mathcal{N}-1$ conditions
\begin{eqnarray}
\frac{\varphi_1^c}{\varphi_1^*}=\left(\frac{\varphi_i^c}{\varphi_i^*}\right)^{\mu_i^2}\,.\label{cond4}
\end{eqnarray}
It should be noted that this is not an attractor: initial conditions for the fields influence the dynamics at all times. As a consequence, a thorough study of initial conditions should be performed whenever a multi-field model with separable quadratic potentials is proposed as a serious scenario of the early universe. However, we are not interested in testing a particular model right now, but in developing intuition by studying a few concrete examples. Naturally these models are chosen such that they are easy to treat. Therefore, we restrict ourselves in most cases to equal energy initial conditions, that is we impose $V_i^*=V_j^*$, which  will simplify matters considerably. Of course this results in the $\mathcal{N}-1$ initial conditions
\begin{eqnarray}
\varphi_i^*=\mu_i\varphi_1^*\,.\label{cond3}
\end{eqnarray}
Next, consider the number of e-foldings defined in (\ref{defN}): this expression integrates to
\begin{eqnarray}
4N=\sum_{i=1}^{\mathcal{N}}\left[ (\varphi_i^{*})^2-(\varphi_i^{c})^2\right]\,.\label{cond2}
\end{eqnarray}
The last missing equation is due to the fact that slow roll ends for one of the fields at $t_c$, that is either one of the slow roll parameters $\eta_i$, $\varepsilon_i$ defined in (\ref{srparameters}) or $\varepsilon$ becomes of order one. We can easily compute the ratio of the $\eta_i$'s, yielding $\eta_i>\eta_j$ if $i>j$. Similarly, the ratio of $\eta_i$ to $\varepsilon_i$ becomes $W/V_i$, which is larger than one so that $\eta_i>\varepsilon_i$. Consequently, the field with the largest mass, that is the $\mathcal{N}$'th field, will leave slow roll first when
\begin{eqnarray}
\eta_{\mathcal{N}}^c=1\,, \label{cond1}
\end{eqnarray}
given that $\varepsilon^c<1$, which will be satisfied in all cases we study.

Equations (\ref{cond4})-(\ref{cond1}) are the $2\mathcal{N}$ conditions which are needed in order to evaluate all fields at $t_*$ and $t_c$. For any concrete models we can solve the above conditions and evaluate $F$. For simplicity,  we suppress the superscript $c$ in the following. 

\subsection{Two fields: $m_1^2/m_2^2=1/2$ \label{case1}}
We start with this case, since it is possible to solve (\ref{cond4})-(\ref{cond1}) without any approximations. Equation (\ref{cond1}) immediately yields
\begin{eqnarray}
\varphi_1^2=2(2-\varphi_2^2)\,,
\end{eqnarray}
and (\ref{cond3}) becomes
\begin{eqnarray}
\varphi_1^{*2}=2\varphi_2^{*2}\,.
\end{eqnarray}
Using these in (\ref{cond2}) yields
\begin{eqnarray}
\varphi_1^{*2}=2\frac{4(N+1)-\varphi_2^2}{3}\,,
\end{eqnarray}
which in turn can be plugged  into (\ref{cond4}) with the result
\begin{eqnarray}
9\frac{(2-\varphi_2^2)^2}{4(N+1)-\varphi_2^2} = 3\varphi_2^2\,.
\end{eqnarray}
This quadratic equation in  $\varphi_2^2$ can be solved to
\begin{eqnarray}
\varphi_2^2&=&2+\frac{N}{2}-\frac{1}{2}\sqrt{4+8N+N^2}\\
&\approx&\frac{3}{N}-\frac{12}{N^2}+\frac{57}{N^3}+\mathcal{O}(N^{-4})\,,
\end{eqnarray} 
where we expanded in terms of $1/N$ in the last step -- we need to keep terms up to order $N^{-3}$, to recapture properly the leading order contribution to $F$. The remaining fields become 
\begin{eqnarray}
\varphi_1^2=4-\frac{6}{N}+\frac{24}{N^2}-\frac{114}{N^3}+\mathcal{O}(N^{-4})\,,\\
\varphi_1^{*2}=\frac{8}{3}(N+1)-\frac{2}{N}+\frac{8}{N^2}-\frac{38}{N^3}+\mathcal{O}(N^{-4})\,,\\
\varphi_2^{*2}=\frac{4}{3}(N+1)-\frac{1}{N}+\frac{4}{N^2}-\frac{19}{N^3}+\mathcal{O}(N^{-4})\,.
\end{eqnarray}
It is now straightforward but somewhat tedious to evaluate all slow roll parameters as well as $u_1$ and $u_2$. In the end, we arrive at
\begin{eqnarray}
F=-\frac{9}{16 N^4}+\mathcal{O}(N^{-5})\,.
\end{eqnarray}

The important feature is the obvious slow roll suppression $F\sim N^{-2\cdot 2}$. This suppression is due to the cancellations that occur within the summation over the $\mathcal{A}$-matrix. As a result, $\mathcal A$ defined in (\ref{defA2field}) becomes proportional to $N^{-4}$, so that even though the prefactor in (\ref{twofieldf}) is of order one, the final result is heavily slow roll suppressed. 
 
\subsection{Two fields: $m_1^2/m_2^2=1/\alpha $ \label{case2}}
This case can still be solved approximately and we will be able to see how $F$, and with that the non-Gaussianity, depends on the ratio of the masses. Taking analogous steps to the discussion in the previous section we arrive at
\begin{eqnarray}
\varphi_1^2&=&\alpha(2-\varphi_2^2)\,,\\
\varphi_1^{*2}&=&\alpha\varphi_2^{*2}\,, \label{eqeninco}\\
\varphi_1^{*2}&=&\frac{4N+2\alpha+\varphi_2^2(1-\alpha)}{1+\frac{1}{\alpha}}\,.
\end{eqnarray}
Using these, and writing $x=\varphi_2^2$ gives
\begin{eqnarray}
\frac{(2-x)^\alpha(1+\alpha)^\alpha}{\left(4N+2\alpha+x(1-\alpha)\right)^{\alpha-1}}=x(1+\alpha)\,,
\end{eqnarray}
which could be solved numerically. However, we refrain from doing so since we are primarily interested in the analytic form of the slow roll suppression of $F$, which can be computed if we make some minor approximations. First, assume that $\varphi_i\ll\varphi_j^*$ with the consequence that (\ref{cond3}) and (\ref{cond2}) immediately yield
\begin{eqnarray}
\varphi_1^{*2}&\approx&\alpha\frac{4N}{1+\alpha}\,,\\
\varphi_2^{*2}&\approx&\frac{4N}{1+\alpha}\,.\\
\end{eqnarray} 
Equation (\ref{cond1}) still leads to
\begin{eqnarray}
\varphi_1^2&=&\alpha(2-\varphi_2^2)\,,\\
\end{eqnarray}
and (\ref{cond4}) becomes 
\begin{eqnarray}
\frac{(2-x)^\alpha}{x}\approx\left(\frac{4N}{\alpha+1}\right)^{\alpha-1}\,.
\end{eqnarray}
If we now Taylor expand the left hand side for $x\ll1$, we arrive at 
\begin{eqnarray}
2^\alpha\left(\frac{1}{x}-\frac{\alpha}{2}+\frac{\alpha(\alpha-1)}{8}x\right)\approx\left(\frac{4N}{\alpha+1}\right)^{\alpha-1}\,,
\end{eqnarray}
where we truncated the expansion such that a quadratic equation for $x$ results. Note that the above expression yields the exact result in case of $\alpha=2$. Solving this equation results in 
\begin{eqnarray}
\varphi_2^2\approx\frac{2^{\alpha-1} \alpha+\left[\frac{4N}{\alpha+1}\right]^{\alpha-1}-\sqrt{4^{\alpha-1}\alpha(2-\alpha)+\left[\frac{4N}{\alpha+1}\right]^{\alpha-1}2^{\alpha}\alpha+\left[\frac{4N}{\alpha+1}\right]^{2\alpha-2}}}{2^{\alpha-2}\alpha(\alpha-1)}\,.
\end{eqnarray}
As in the previous section, we can now expand the field values to any desired order in $1/N$, compute the slow roll parameters, evaluate $u_1$ and $u_2$ and in the end compute the non-Gaussianity due to $F$. Doing so we arrive at 
\begin{eqnarray}
F\sim\frac{1}{N^{2\alpha}}\,,\label{sclaingalp}
\end{eqnarray} 
for $\alpha=2,3,4,...$, and $F=0$ for $\alpha=1$.
 The suppression is again due to cancellations within the $\mathcal{A}$-matrix. Even though the proportionality factor increases with $\alpha$, it is easily dominated by the additional suppression due to $N^{-2\alpha}$. One should note that our assumption $\varphi_i\ll\varphi_j^*$ breaks down for $\alpha\approx 5$, since $\varphi_1^2 \approx \varphi_2^{*2}/4$ already, if we use $N=60$. 

We conclude that non-Gaussianities become  more increasingly suppressed as the mass difference increases. If the masses are similar ($\alpha\approx 1$), we should still have a suppression of $1/N^2$. Of course the non-Gaussianity due to the $\mathcal{A}$-matrix vanishes identically if $\alpha=1$, that is if the masses are equal.  Hence, we predict the largest non-Gaussianity to occur for nearly equal masses. To be specific, if we write $m_1^2/m_2^2=1-\delta$ with $\delta\ll1$, we expect $F\sim \delta^\beta/N^2$ with some exponent $\beta$. As soon as $\delta$ becomes of order one, the additional slow roll suppression kicks in and suppresses the signal.

One assumption we made was the equal energy initial condition (\ref{eqeninco}), and one might wonder how a different choice effects $F$: it turns out that only the numerical prefactor in (\ref{sclaingalp}) changes if we choose different initial values for the fields (e.g. for $\alpha=2$ it changes from $-9/16$ to $-4$ if $\varphi_1^*=\varphi_2^*$ is used), but the important suppression factor $N^{-2\alpha}$ remains unaltered. As a consequence, we will restrain ourselves to the equal energy initial condition in the next section, where we will be able to derive the exponent $\beta$ and the proportionality factor in general for multi-field models. 

\subsection{$\mathcal{N}$ fields: $m_1^2/m_i^2=1-\delta_i$ \label{case3}}
Let us now consider the case of $\mathcal{N}$ fields and focus on a narrow mass spectrum, since a broad distribution is expected to result in heavily  suppressed non-Gaussianities, based on the analysis in the previous section. Another reason to focus on narrow mass spectra is that broad distributions are not well suited for assisted inflation, since the heavy fields would roll quickly to their minimum without contributing much to inflation. 

Therefore consider 
\begin{eqnarray}
\frac{m_1^2}{m_i^2}&\equiv&1-\delta_i\,, 
\end{eqnarray}
with $\delta_i\ll1$ and  $\delta_i>\delta_j$ if $i>j$, that is we order the fields according to their masses. We characterize the width of the mass distribution by
\begin{eqnarray}
\delta\equiv\frac{1}{\mathcal N}\sum_{i=1}^{\mathcal N}\delta_i
\end{eqnarray}
and 
\begin{eqnarray}
\tilde{\delta}^2\equiv\frac{1}{\mathcal N}\sum_{i=1}^{\mathcal N}\delta_i^2\,.
\end{eqnarray}

Our first task is again to find the field values at $t_*$ and $t_c$ via (\ref{cond4})-(\ref{cond1}). What simplifies matters is that we can expand in terms of the $\delta_i$. However, as we shall see later on, terms linear in $\delta_i$ cancel out exactly in the expression for $F$. Therefore we keep all terms up to $\delta_i^2$ right from the start.

By using (\ref{cond4}) and (\ref{cond3}) in (\ref{cond1}) we get
\begin{eqnarray}
\frac{2}{1-\delta_{\mathcal{N}}}=\varphi_1^{*2}\sum_{i=1}^{\mathcal{N}}\left(\frac{\varphi_1^2}{\varphi_1^{*2}}\right)^{\frac{1}{1-\delta_i}}\,,
\end{eqnarray}
and (\ref{cond2}) turns into 
\begin{eqnarray}
4N=\varphi_1^{*2}\sum_{i=1}^{\mathcal{N}}(1-\delta_i)-\varphi_1^{*2}\sum_{i=1}^{\mathcal{N}}(1-\delta_i)\left(\frac{\varphi_1^2}{\varphi_1^{*2}}\right)^{\frac{1}{1-\delta_i}}\,,
\end{eqnarray}
without expanding anything so far. Next, noting that
\begin{eqnarray}
\left(\frac{\varphi_1^2}{\varphi_1^{*2}}\right)^{\frac{1}{1-\delta_i}}\approx \frac{\varphi_1^2}{\varphi_1^{*2}}
\left(1+(\delta_i+\delta_i^2)\ln\left(\frac{\varphi_1^2}{\varphi_1^{*2}}\right)+\delta_i^2\frac{1}{2}\ln^2\left(\frac{\varphi_1^2}{\varphi_1^{*2}}\right)+\mathcal{O}(\delta_i^3)\right)
\end{eqnarray}
and using the definitions of $\delta$ and $\tilde{\delta}^2$, we arrive at
\begin{eqnarray}
\nonumber \frac{\mathcal{N}}{2}\varphi_1^{*2}&\approx&(2N+1)+\delta_{\mathcal{N}}+\delta\left(2N+1-\varphi_1^2\frac{\mathcal{N}}{2}\right)\\
&&+\delta_{\mathcal{N}}^2-\varphi_1^2\frac{\mathcal{N}}{2}\left(\delta^2+\tilde{\delta}^2\ln\left(\frac{\varphi_1^2}{\varphi_1^{*2}}\right)\right)+\delta\delta_{\mathcal{N}}+\delta^2(2N+1)\,,\\
\nonumber \frac{\mathcal{N}}{2}\varphi_1^{2}&\approx&1+\delta_{\mathcal{N}}-\delta\ln\left(\frac{\varphi_1^2}{\varphi_1^{*2}}\right)
\\
&&+\delta_{\mathcal{N}}^2-(\tilde{\delta}^2+\delta_{\mathcal N}\delta)\ln\left(\frac{\varphi_1^2}{\varphi_1^{*2}}\right)+\left(\delta^2-\frac{\tilde{\delta}^2}{2}\right)\ln^2\left(\frac{\varphi_1^2}{\varphi_1^{*2}}\right)\,,
\end{eqnarray}
where we kept all terms up to second order. To solve these coupled equations, we iteratively insert them into each other until we have all terms up to second order. The solution reads
\begin{eqnarray}
 \frac{\mathcal{N}}{2}\varphi_1^{*2}&\approx&2N+1+\delta_{\mathcal{N}}+2N \delta+(\tilde{\delta}^2-\delta^2)\gamma+\delta_{\mathcal{N}}^2+2N\delta^2 \,,\\
 \frac{\mathcal{N}}{2}\varphi_1^{2}&\approx&1+\delta_{\mathcal{N}}+\gamma\delta+\frac{2N}{2N+1}\delta(\delta-\delta_{\mathcal{N}})\nonumber\\
&&+\delta_{\mathcal{N}}^2+(\tilde{\delta}^2-\delta^2+\delta\delta_{\mathcal{N}})\gamma+\frac{2\delta^2-\tilde{\delta}^2}{2}\gamma^2\,,
\end{eqnarray}
where we introduced
\begin{eqnarray}
\gamma\equiv\ln(2N+1)\,,
\end{eqnarray}
which is of order one. All other fields follow directly from (\ref{cond4}) and (\ref{cond3}), which have to be properly expanded too.

Now that we have the field values, we are ready to compute the slow roll parameters. The easiest one is $\eta_i=m_i^2/m_{\mathcal{N}}^2$,  which becomes
\begin{eqnarray}
\eta_i&=&\frac{1-\delta_{\mathcal{N}}}{1-\delta_i}\\
 &\approx&1-\delta_{\mathcal{N}}+\delta_i-\delta_{\mathcal{N}}\delta_i+\delta_i^2\,.
\end{eqnarray} 
Next, we would like to compute $\varepsilon=\sum\varepsilon_i$ since the ratio of $\eta_i$ and $\varepsilon$ appears in the $\mathcal{A}$-matrix. Noting that $\varepsilon_i=\eta_iV_i/W$ as well as $W=m_{\mathcal{N}}^2$, and plugging in $\varphi_i^2$ by using (\ref{cond4}) and (\ref{cond3}) with $\varphi_1^2$ and $\varphi_1^{*2}$ from above we arrive after some tedious but straightforward algebra at
\begin{eqnarray}
\varepsilon\approx1-\delta_{\mathcal{N}}+\delta-\delta\delta_{\mathcal{N}}+\tilde{\delta}^2(\gamma+1)+\delta^2\gamma\,.\label{toteps}
\end{eqnarray}
Note that $\varepsilon<1$ since $\delta_N>\delta$, so that $\eta_{\mathcal{N}}$ is indeed the largest slow roll parameter. The ratio we are interested in becomes
\begin{eqnarray}
1-\frac{\eta_i}{\varepsilon}\approx\delta-\delta_i-\delta_i^2+\delta_i\delta+\tilde{\delta}^2(\gamma+1)-\delta^2(1-\gamma)\,.\label{etaexp}
\end{eqnarray}

At this point we should step back for a second and have a look at the expression for $F$: first, note that $\sum_i(\delta-\delta_i)=0$. Because of this and since the leading order contribution of $1-\eta_i/\varepsilon$ is already first order in delta,  we  know that $F$ is identical to zero up to first order in delta. This is the reason why we computed $1-\eta_i/\varepsilon$ up to second order. Next, since $1-\eta_i/\varepsilon$ has no zeroth order contribution, we only need to evaluate the remaining constituents of $F$ up to first order in delta, which simplifies matters quite a bit. 

Doing so leads after a lot more straightforward but tedious algebra to
\begin{eqnarray}
\nonumber F&\approx&\frac{\tilde{\delta}^2-\delta^2}{(2N+1)^2}\left(1-\frac{2}{\mathcal{N}(2N+1)}-\frac{1}{\mathcal{N}^2}\left(1-\frac{2}{2N+1}\right)\right)\\
&&-\frac{2\tilde{\delta}^2\ln(2N+1)}{(2N+1)^2}\left(1-\frac{2}{\mathcal{N}}+\frac{1}{\mathcal{N}^2}\right) \label{Fresult}
\end{eqnarray}
This is our final result, but before we discuss its implications, it might be instructive to see how at least the leading order contribution in the number of fields comes about. By looking at the general expression for $F$ in (\ref{sum}) as well as the definition of the $\mathcal{A}$-matrix in (\ref{defA}), one can convince oneself that
\begin{eqnarray}
\nonumber F\big|_{\mathcal{O}(\mathcal{N}^0)}&=&
-\frac{W^2}{W^{2}_*}\Big|_{\mathcal{O}(\delta^0)}\left(\sum_{k,l=1}^{\mathcal{N}}\frac{u_k^2}{\varepsilon_k^*}\Big|_{\mathcal{O}(\delta^0)}\right)^{-2}
\sum_{k,l=1}^{\mathcal{N}}
\frac{u_k u_l}{\varepsilon_k^*\varepsilon_l^*}
\Big|_{\mathcal{O}(\delta^0)}\\
\nonumber &&\times\Bigg( \sum_{j=1}^{\mathcal{N}}\left(\varepsilon_j\big|_{\mathcal{O}(\delta)}-\varepsilon_j\big|_{\mathcal{O}(\delta^0)}\right) \frac{\varepsilon_k\varepsilon_l}{\varepsilon^2}\Big|_{\mathcal{O}(\delta^0)}\left(1-\frac{\eta_j}{\varepsilon}\right)\Big|_{\mathcal{O}(\delta)} \\
&&+\sum_{j=1}^{\mathcal{N}}\left(\varepsilon_j\frac{\varepsilon_k\varepsilon_l}{\varepsilon^2}\right)\Big|_{\mathcal{O}(\delta^0)}\left(-\frac{\eta_j}{\varepsilon}\Big|_{\mathcal{O}(\delta^2)}+\frac{\eta_j}{\varepsilon}\Big|_{\mathcal{O}(\delta)}\right) \Bigg)+\mathcal{O}(\delta^3)\,,
\end{eqnarray}
where $|_{\mathcal{O}(\delta^{\beta})}$ means that the adjacent quantity has to be expanded up to order $\beta$ in all deltas. To evaluate this expression we use $1-\eta_j/\varepsilon$ from (\ref{etaexp}), $\varepsilon$ from (\ref{toteps}) and 
\begin{eqnarray}
\varepsilon_k\Big|_{\mathcal{O}(\delta^1)}&=&\frac{1}{\mathcal{N}}\left(1-\delta_k(\gamma-1)-\delta_{\mathcal{N}}+\gamma\delta\right)\,,\\
\frac{u_k}{\varepsilon_k^*}\Big|_{\mathcal{O}(\delta^0)}&=&2N+1\,,\\
\frac{u_k^2}{\varepsilon_k^*}\Big|_{\mathcal{O}(\delta^0)}&=&\frac{2N+1}{\mathcal{N}}\,,\\
\frac{W^2}{W^{2}_*}\Big|_{\mathcal{O}(\delta^0)}&=&\frac{1}{(2N+1)^2}\,,
\end{eqnarray}
which leads to the $\mathcal{O}(\mathcal{N}^0)$ contribution to $F$ from above.

There are two interesting features to our result (\ref{Fresult}): first, the expression is again suppressed by the number of e-foldings ($\propto N^{-2}$), just as expected from our experience with the two-field cases. Next, the leading order contribution is not enhanced by the number of fields, as one might naively expect, but is of order $\mathcal{O}(\delta^2)$. As a consequence, the contribution to the non-Gaussianity due to the $\mathcal{A}$-matrix is negligible, even for multiple fields. 

We restricted ourselves to a narrow mass spectrum here -- however, for a broader mass spectrum we expect an even stronger suppression by inverse powers of the number of e-foldings, based on our experience with two-field models.

\subsection{Discussion}

We saw in the previous three subsections that there is little hope for a considerable amount of non-Gaussianity due to the evolution of modes once they cross the horizon in multi-field models of inflation. We restricted ourselves to separable potentials when we derive the expression for the nonlinear parameter in (\ref{f_NL}), and looked at a few specific models with potentials of the $m^2\varphi^2$ type: in the case of two fields,  the contribution due to the $\mathcal{A}$-matrix is suppressed by the number of e-foldings, with an exponent given by twice the ratio of the heavier square mass to the lighter one.

Extrapolating this result to the multi-field case, we concluded that a narrow spectrum would be the most promising candidate for a large non-Gaussian signal. This, and the fact that the most useful multi-field models have a narrow spectrum, eg. in the case of assisted inflation, lead us to evaluate the non-Gaussianity in a general multi-field scenario with a narrow mass distribution. As expected, we arrived at an expression that is suppressed by the number of e-foldings, in agreement with our experience from the two-field case. What is more, no enhancement due to the potentially large number of fields was found. Finally, the expression scaled like $\delta^2$, where $\delta$ is a measure of the mass distributions width (the larger $\delta$, the broader the spectrum).

For the $\mathcal{N}$-field case, we used equal energy initial conditions only, for reasons of simplicity and since we saw in the two field model that only the numerical prefactor gets altered by a different choice. Consequently, we expect the general scaling behavior and the suppression by the number of e-foldings to be generic features, independent of the chosen initial field values. Nevertheless, one should perform a more careful analysis of initial conditions for any concrete model of the early universe, if the model is based on multiple fields with quadratic potentials (like N-flation). The machinery for evaluating a non-Gaussian signal developed in this paper can of course be used.

We did not consider more intricate potentials in our case studies, like $\lambda\varphi^4$ potentials, or broad mass spectra, such as in N-flation, since our main aim here was to develop the general formalism and to demonstrate its applicability. We do not expect drastic new features for different potentials, but a thorough analysis of models in the literature should be performed, including the computation of the first term in (\ref{f_NL}) \cite{inprep}.

It should be noted that we have restricted ourselves to slowly rolling fields throughout. However, if the mass spectrum is broad, fields will leave slow roll early on, while inflation still commences. These fields  will lead to an additional production of non-Gaussianities, due to the conversion of isocurvature modes to adiabatic ones \cite{Rigopoulos:2005us}; based on the simple models studied in the literature we expect these signals to be transient \cite{Rigopoulos:2005us}, but more intricate models should be examined carefully before drawing general conclusions. We also neglected any cross coupling terms between fields, which might be yet another source of non-Gaussianity, see e.g. \cite{Bernardeau:2006tf}.

Before we conclude,  we quickly derive the general expression for the scalar spectral index in case of a separable potential, that includes the evolution of modes once they cross the horizon. Since this expression incorporates the effect of isocurvature modes, it might be quite useful whenever these modes are present, e.g.~in models incorporating compactifications of higher dimensions.

\section{Scalar Spectral Index \label{scind}}
With our knowledge of $N_{,k}$ and ${N_{,kl}}$ from  section \ref{nongaussianities} we can easily evaluate the scalar spectral index in the $\delta N$ formalism.  
In \cite{Vernizzi:2006ve} the general expression for $n_\zeta$ was derived as 
\begin{eqnarray}
n_\zeta-1=-2\varepsilon^*+\frac{2}{H_*}\frac{\sum_{k,l=1}^{\mathcal{N}} \dot{\varphi}_k^*N_{,kl}N_{,l}}{\sum_{k=1}^{\mathcal{N}}N_{,k}^2}\,,
\end{eqnarray}
which can be shown to be equivalent with the result of Sasaki and Stewart \cite{Sasaki:1995aw} within the slow roll regime. If we insert our expressions for the derivatives of the expansion rate from (\ref{N,k}) and (\ref{N,kl}) we arrive at
\begin{eqnarray}
n_\zeta-1=-2\varepsilon^*-4\frac{\sum_{k=1}^{\mathcal{N}}u_k\left(1-\frac{\eta_k^*}{2\varepsilon_k^*}u_k\right)+\sum_{k,l=1}^{\mathcal{N}}\frac{u_k}{\varepsilon_k^*}\mathcal{A}_{kl}}{\sum_{k=1}^{\mathcal{N}}\frac{u_k^{2}}{\varepsilon_k^*}}\,,
\end{eqnarray} 
where we made use of the definition $\sqrt{2\varepsilon_k^*}=W_*3H_*\dot{\varphi}^*$ and the Friedman equation $3H^{2}_*=W_*$.
It is easy to see that the last sum in the numerator vanishes, since we established already  in (\ref{rowA}) that the sum over a row or column of the $\mathcal{A}$-matrix vanishes. Henceforth, the scalar spectral index simplifies to
\begin{eqnarray}
n_\zeta-1=-2\varepsilon^*-4\frac{\sum_{k=1}^{\mathcal{N}}u_k\left(1-\frac{\eta_k^*}{2\varepsilon_k^*}u_k\right)}{\sum_{k=1}^{\mathcal{N}}\frac{u_k^{2}}{\varepsilon_k^*}}\,.\label{scalind}
\end{eqnarray} 
This result reduces to the one of \cite{Vernizzi:2006ve} in the case of two fields. It is interesting to note that the $\mathcal{A}$-matrix, which was due to $\partial Z_k^c/\partial\varphi^*_l$, cancels out of the expression for $n_\zeta-1$. Obviously, this matrix is not constrained by observations of a nearly scale invariant spectral index. 

This expression includes the effect of possible isocurvature modes and might be quite useful for models where such modes arise. We postpone a detailed study of ({\ref{scalind}})'s application to concrete models of the early universe to a forthcoming publication.

\section{Conclusions \label{sec:con}} 
In this article, we derived the general expression of the nonlinear parameter for separable potentials, within the slow roll approximation and including the evolution of perturbations once they cross the horizon. 

This formalism was then applied to several specific models with quadratic potentials. We find that non-Gaussianities are suppressed by  the volume expansion rate. The power of the rate is twice the ratio of the heavier mass squared to the lighter one squared in two-field models. 

Extrapolating this result to multiple fields, we expect the suppression to be stronger for broad spectra than for narrow spectra. As a consequence, we focused on narrow spectra only for the general multi-field case. We recover the expected quadratic suppression with respect to the volume expansion rate, but we are also able to derive the explicit expression for the nonlinear parameter in terms of the width of the spectrum.

Based on these case studies, we expect that multi-field models of inflation which are well described by the slow roll approximation cannot generate a large non-Gaussian signal. Consequently, we cannot use $f_{NL}$ to discriminate between multi-field models of assisted inflation and their single field analogs. Nevertheless, given a concrete multi-field model one should compute its non-Gaussianity, to verify that the general expectation is indeed true for the model at hand, since a dependence on the mass spectrum and, in case of quadratic potentials, the initial field values is present. Furthermore, we did not examine more intricate potentials, such as quartic or exponential ones, which clearly warrant further study \cite{inprep}.

  As a bonus, we also computed the general expression of the scalar spectral index, again including the evolution of perturbations once they cross the horizon. This result should be applicable to computations of  the effects of isocurvature modes. We postpone an application of this result to a forthcoming publication.

%%%%%%%%%%%%%%%%%%%%%%%%%%%%%%%%%%%%
\begin{acknowledgments}

We would like to thank Diana Battefeld, Robert Brandenberger and Eugene Lim for comments on the draft. T.B. would like to acknowledge the hospitality of Yale University where this work was initiated and completed.  We are also thankful for useful conversations on this topic with Daniel Baumann, Andrew Liddle and Liam McAllister.    RE is supported in part by the United States Department of Energy, grant DE-FG02-92ER-40704.  

\end{acknowledgments}
%%%%%%%%%%%%%%%%%%%%%%%%%%%%%%%%%%%%


\begin{thebibliography}{}

%\cite{Linde:1991km}
\bibitem{Linde:1991km}
  A.~D.~Linde,
  ``Axions in inflationary cosmology,''
  Phys.\ Lett.\ B {\bf 259}, 38 (1991).
  %%CITATION = PHLTA,B259,38;%%

%\cite{Linde:1993cn}
\bibitem{Linde:1993cn}
  A.~D.~Linde,
  ``Hybrid inflation,''
  Phys.\ Rev.\ D {\bf 49}, 748 (1994)
  [arXiv:astro-ph/9307002].
  %%CITATION = ASTRO-PH 9307002;%%


%\cite{Copeland:1994vg}
\bibitem{Copeland:1994vg}
  E.~J.~Copeland, A.~R.~Liddle, D.~H.~Lyth, E.~D.~Stewart and D.~Wands,
  ``False vacuum inflation with Einstein gravity,''
  Phys.\ Rev.\ D {\bf 49}, 6410 (1994)
  [arXiv:astro-ph/9401011].
  %%CITATION = ASTRO-PH 9401011;%%

%\cite{Liddle:1998jc}
\bibitem{Liddle:1998jc}
  A.~R.~Liddle, A.~Mazumdar and F.~E.~Schunck,
  ``Assisted inflation,''
  Phys.\ Rev.\ D {\bf 58}, 061301 (1998)
  [arXiv:astro-ph/9804177].
  %%CITATION = ASTRO-PH 9804177;%%

%\cite{Kanti:1999vt}
\bibitem{Kanti:1999vt}
  P.~Kanti and K.~A.~Olive,
  ``On the realization of assisted inflation,''
  Phys.\ Rev.\ D {\bf 60}, 043502 (1999)
  [arXiv:hep-ph/9903524].
  %%CITATION = HEP-PH 9903524;%%

%\cite{Dimopoulos:2005ac}
\bibitem{Dimopoulos:2005ac}
  S.~Dimopoulos, S.~Kachru, J.~McGreevy and J.~G.~Wacker,
  ``N-flation,''
  arXiv:hep-th/0507205.
  %%CITATION = HEP-TH 0507205;%%

%\cite{Easther:2005zr}
\bibitem{Easther:2005zr}
  R.~Easther and L.~McAllister,
  ``Random matrices and the spectrum of N-flation,''
  JCAP {\bf 0605}, 018 (2006)
  [arXiv:hep-th/0512102].
  %%CITATION = HEP-TH 0512102;%%

%\cite{Becker:2005sg}
\bibitem{Becker:2005sg}
  K.~Becker, M.~Becker and A.~Krause,
  %``M-theory inflation from multi M5-brane dynamics,''
  Nucl.\ Phys.\ B {\bf 715}, 349 (2005)
  [arXiv:hep-th/0501130].
  %%CITATION = HEP-TH 0501130;%%

%\cite{Maldacena:2002vr}
\bibitem{Maldacena:2002vr}
  J.~M.~Maldacena,
   ``Non-Gaussian features of primordial fluctuations in single field
  inflationary models,''
  JHEP {\bf 0305}, 013 (2003)
  [arXiv:astro-ph/0210603].
  %%CITATION = ASTRO-PH 0210603;%%  
  
  %\cite{Acquaviva:2002ud}
\bibitem{Acquaviva:2002ud}
  V.~Acquaviva, N.~Bartolo, S.~Matarrese and A.~Riotto,
  ``Second-order cosmological perturbations from inflation,''
  Nucl.\ Phys.\ B {\bf 667}, 119 (2003)
  [arXiv:astro-ph/0209156].
  %%CITATION = ASTRO-PH 0209156;%%

%\cite{Creminelli:2003iq}
\bibitem{Creminelli:2003iq}
  P.~Creminelli,
  ``On non-gaussianities in single-field inflation,''
  JCAP {\bf 0310}, 003 (2003)
  [arXiv:astro-ph/0306122].
  %%CITATION = ASTRO-PH 0306122;%%

%\cite{Babich:2004gb}
\bibitem{Babich:2004gb}
  D.~Babich, P.~Creminelli and M.~Zaldarriaga,
  ``The shape of non-Gaussianities,''
  JCAP {\bf 0408}, 009 (2004)
  [arXiv:astro-ph/0405356].
  %%CITATION = ASTRO-PH 0405356;%%
  
  
  %\cite{Seery:2005wm}
\bibitem{Seery:2005wm}
  D.~Seery and J.~E.~Lidsey,
  ``Primordial non-gaussianities in single field inflation,''
  JCAP {\bf 0506}, 003 (2005)
  [arXiv:astro-ph/0503692].
  %%CITATION = ASTRO-PH 0503692;%%

%\cite{Barnaby:2006km}
\bibitem{Barnaby:2006km}
  N.~Barnaby and J.~M.~Cline,
  %``Nongaussianity from Tachyonic Preheating in Hybrid Inflation,''
  arXiv:astro-ph/0611750.
  %%CITATION = ASTRO-PH 0611750;%%

%\cite{Enqvist:2004ey}
\bibitem{Enqvist:2004ey}
  K.~Enqvist, A.~Jokinen, A.~Mazumdar, T.~Multamaki and A.~Vaihkonen,
  %``Non-Gaussianity from Preheating,''
  Phys.\ Rev.\ Lett.\  {\bf 94}, 161301 (2005)
  [arXiv:astro-ph/0411394].
  %%CITATION = ASTRO-PH 0411394;%%
  
  %\cite{Vernizzi:2006ve}
\bibitem{Vernizzi:2006ve}
  F.~Vernizzi and D.~Wands,
  ``Non-Gaussianities in two-field inflation,''
  JCAP {\bf 0605}, 019 (2006)
  [arXiv:astro-ph/0603799].
  %%CITATION = ASTRO-PH 0603799;%%

%\cite{Seery:2005gb}
\bibitem{Seery:2005gb}
  D.~Seery and J.~E.~Lidsey,
  ``Primordial non-gaussianities from multiple-field inflation,''
  JCAP {\bf 0509}, 011 (2005)
  [arXiv:astro-ph/0506056].
  %%CITATION = ASTRO-PH 0506056;%%

  %\cite{Kim:2006te}
\bibitem{Kim:2006te}
  S.~A.~Kim and A.~R.~Liddle,
  ``Nflation: Non-gaussianity in the horizon-crossing approximation,''
  arXiv:astro-ph/0608186.
  %%CITATION = ASTRO-PH 0608186;%%

\bibitem{Starobinski}
A.~A.~Starobinsky, JETP Lett. {\bf 42}, 152 (1985) [Pis. Hz. Esp.
Tor. Fizz. {\rm 42}, 124 (1985)].

%\cite{Sasaki:1995aw}
\bibitem{Sasaki:1995aw}
  M.~Sasaki and E.~D.~Stewart,
   ``A General Analytic Formula For The Spectral Index Of The Density
  Perturbations Produced During Inflation,''
  Prog.\ Theor.\ Phys.\  {\bf 95}, 71 (1996)
  [arXiv:astro-ph/9507001].
  %%CITATION = ASTRO-PH 9507001;%%

%\cite{Lyth:2005fi}
\bibitem{Lyth:2005fi}
  D.~H.~Lyth and Y.~Rodriguez,
  ``The inflationary prediction for primordial non-gaussianity,''
  Phys.\ Rev.\ Lett.\  {\bf 95}, 121302 (2005)
  [arXiv:astro-ph/0504045].
  %%CITATION = ASTRO-PH 0504045;%%

%\cite{Alabidi:2005qi}
\bibitem{Alabidi:2005qi}
  L.~Alabidi and D.~H.~Lyth,
  ``Inflation models and observation,''
  JCAP {\bf 0605}, 016 (2006)
  [arXiv:astro-ph/0510441].
  %%CITATION = ASTRO-PH 0510441;%%

%\cite{Rigopoulos:2005us}
\bibitem{Rigopoulos:2005us}
  G.~I.~Rigopoulos, E.~P.~S.~Shellard and B.~W.~van Tent,
  ``Non-Gaussianity as a new observable in multifield inflation?,''
  arXiv:astro-ph/0511041.
  %%CITATION = ASTRO-PH 0511041;%%

  %\cite{Chen:2006nt}
\bibitem{Chen:2006nt}
  X.~Chen, M.~x.~Huang, S.~Kachru and G.~Shiu,
   ``Observational signatures and non-Gaussianities of general single field inflation,''
  arXiv:hep-th/0605045.
  %%CITATION = HEP-TH 0605045;%%

%\cite{Lyth:2003im}
\bibitem{Lyth:2003im}
  D.~H.~Lyth and D.~Wands,
  ``Conserved cosmological perturbations,''
  Phys.\ Rev.\  D {\bf 68}, 103515 (2003)
  [arXiv:astro-ph/0306498].
  %%CITATION = PHRVA,D68,103515;%%

%\cite{Rigopoulos:2003ak}
\bibitem{Rigopoulos:2003ak}
  G.~I.~Rigopoulos and E.~P.~S.~Shellard,
  ``The separate universe approach and the evolution of nonlinear  superhorizon
  cosmological perturbations,''
  Phys.\ Rev.\  D {\bf 68}, 123518 (2003)
  [arXiv:astro-ph/0306620].
  %%CITATION = PHRVA,D68,123518;%%

%\cite{Byrnes:2006vq}
\bibitem{Byrnes:2006vq}
  C.~T.~Byrnes, M.~Sasaki and D.~Wands,
  ``The primordial trispectrum from inflation,''
  Phys.\ Rev.\  D {\bf 74}, 123519 (2006)
  [arXiv:astro-ph/0611075].
  %%CITATION = PHRVA,D74,123519;%%
  

%\cite{Spergel:2006hy}
\bibitem{Spergel:2006hy}
  D.~N.~Spergel {\it et al.},
   ``Wilkinson Microwave Anisotropy Probe (WMAP) three year results:
  Implications for cosmology,''
  arXiv:astro-ph/0603449.
  %%CITATION = ASTRO-PH 0603449;%%

%\cite{Easther:2005nh}
\bibitem{Easther:2005nh}
  R.~Easther and J.~T.~Giblin,
  ``The Hubble slow roll expansion for multi field inflation,''
  Phys.\ Rev.\ D {\bf 72}, 103505 (2005)
  [arXiv:astro-ph/0505033].
  %%CITATION = ASTRO-PH 0505033;%%

%\cite{Lyth:1998xn}
\bibitem{Lyth:1998xn}
  D.~H.~Lyth and A.~Riotto,
  ``Particle physics models of inflation and the cosmological density
  perturbation,''
  Phys.\ Rept.\  {\bf 314}, 1 (1999)
  [arXiv:hep-ph/9807278].
  %%CITATION = HEP-PH 9807278;%%


\bibitem{inprep}
D.~Battefeld and T.~Battefeld, ``Non-Gaussianities in N-flation,'' 
arXiv:hep-th/0703012.
  %%CITATION = HEP-TH 0703012;%%

%\cite{Bernardeau:2006tf}
\bibitem{Bernardeau:2006tf}
  F.~Bernardeau, T.~Brunier and J.~P.~Uzan,
  %``Models of inflation with primordial non-Gaussianities,''
  AIP Conf.\ Proc.\  {\bf 861}, 821 (2006)
  [arXiv:astro-ph/0604200].
  %%CITATION = APCPC,861,821;%%



%%%%%%%%%%%%%%%%%%%%%%%%%%%%%%%%%%%%
% Belows ref. are not cited in the text yet

%\cite{Kim:2006ys}
%\bibitem{Kim:2006ys}
%  S.~A.~Kim and A.~R.~Liddle,
%  ``Nflation: Multi-field inflationary dynamics and perturbations,''
%  Phys.\ Rev.\ D {\bf 74}, 023513 (2006)
%  [arXiv:astro-ph/0605604].
  %%CITATION = ASTRO-PH 0605604;%%


%\cite{Piao:2006nm}
%\bibitem{Piao:2006nm}
%  Y.~S.~Piao,
%  ``On perturbation spectra of N-flation,''
%  arXiv:gr-qc/0606034.
  %%CITATION = GR-QC 0606034;%%



\end{thebibliography}
\end{document}